# Popularity and Performance: A Large-Scale Study

PETER KRAFFT*, JULIA ZHENG*, and EREZ SHMUELI, Massachusetts Institute of Technology
NICOLÁS DELLA PENNA, Australian National University
JOSHUA TENENBAUM and ALEX PENTLAND, Massachusetts Institute of Technology

## 1. INTRODUCTION

Social scientists have long sought to understand why certain people, items, or options become more popular than others. One seemingly intuitive theory is that inherent value drives popularity. An alternative theory claims that popularity is driven by the rich-get-richer effect of cumulative advantage—certain options become more popular not because they are higher quality but because they are already relatively popular. Realistically, it seems likely that popularity is driven by neither one of these forces alone but rather both together.

Recently researchers have begun using large-scale online experiments to study the effect of cumulative advantage in realistic scenarios [Salganik et al. 2006], [Muchnik et al. 2013], but there have been no large-scale studies of the combination of these two effects. We are interested in studying a case where decision-makers observe explicit signals of both the popularity and the quality of various options. We derive a model for change in popularity as a function of past popularity and past perceived quality. Our model implies that we should expect an interaction between these two forces—popularity should amplify the effect of quality, so that the more popular an option is, the faster we expect it to increase in popularity with better perceived quality. We use a data set from eToro.com, an online social investment platform, to support this hypothesis.

## 2. MODEL

Our model describes the evolution of popularity of an individual action $a$ (e.g., choosing to buy a particular brand or following a particular person). We assume a fixed population of $N$ agents that all have the option to take action $a$ at each of a series of discrete times. That is, at each time every agent decides whether to take action $a$ in that time step. We assume that agents want to make good choices, where the action being $good$ is defined in some suitable domain-specific way. We also assume that at the end of each time step, all agents observe a single new signal of the action's quality, as well as how many other agents took the action in that step. We denote the number of agents taking the action at time $t$ as $n_t$ and the signal of its quality at time $t$ as $q_t$. We assume agents attempt to evaluate whether the action is $good$ using Bayesian inference. In this case users are tasked with computing the posterior distribution $P(good \mid q_1, \ldots, q_t)$. Further we assume that agents choose whether to take the action at a particular time step via probability matching, which means that each agent decides whether to take the action at round $t + 1$ with probability $P(good \mid q_1, \ldots, q_t)$, i.e. agents match the probability that they take action $a$ with the probability that the action is good. This assumption of probabilistic decision-making has precedent in cognitive science [Vul et al. 2009], animal behavior [Prez-Escudero and de Polavieja 2011], and economics [Anderson and Holt 1997].

---

*The first two authors contributed equally.

Then, letting $\alpha$ be an arbitrary positive constant (used as a smoothing parameter), we have

$$P(good \,|\, q_1, \ldots, q_t) = \frac{P(q_t \,|\, good, q_1, \ldots, q_{t-1})}{P(q_t \,|\, q_1, \ldots, q_{t-1})} P(good \,|\, q_1, \ldots, q_{t-1}) \approx \frac{P(q_t \,|\, good, q_1, \ldots, q_{t-1})}{P(q_t \,|\, q_1, \ldots, q_{t-1})} \frac{n_{t-1} + \alpha}{N + \alpha},$$

where the approximation follows from an interesting observation: Since users are probability matching, previous popularity actually approximates the posterior distribution from the last time step, and hence $P(good \,|\, q_1, \ldots, q_{t-1})$ is given by the (smoothed) proportion of agents that chose to take the action in the last time step.

Finally, noting that by the same argument future popularity will estimate $P(good \,|\, q_1, \ldots, q_t)$, assuming $\alpha$ is small relative to $N$, and letting $f(q_1, \ldots, q_t) = \frac{P(q_t \,|\, good, q_1, \ldots, q_{t-1})}{P(q_t \,|\, q_1, \ldots, q_{t-1})}$, we see that the expected change in popularity in the next time step is given by

$$n_{t+1} - n_t \approx (f(q_1, \ldots, q_t) - 1) \cdot n_t + f(q_1, \ldots, q_t) \cdot \alpha.$$

When $f$ is a monotonically increasing function of $q_t$, this equation implies that we should expect a synergy between popularity and quality whereby increasing the popularity of an action should amplify the boost in popularity the action would get from a new signal of high quality.

## 3. DATA

The data set we use to test this hypothesis was provided to us by the eToro company. eToro offers a website that incorporates several different trading platforms alongside a social trading network. Individuals can trade in the commodities, stock, and currencies markets. Furthermore, traders can conduct their own trades or view and copy trades made by other users, all using real money. The data set consists of transactions from June 13, 2011 to November 20, 2013 from their website, eToro.com.

One particularly interesting feature of the site is the ability that users have to mirror other users, automatically copying all of the trades they make. Importantly, users can choose who they want to mirror by viewing various measurements of performance as well as the current popularities of those traders. Thus the number of copiers a user has in the future, i.e. the future popularity of that user, could be influenced both by explicit signals of that user's current popularity and of that user's quality, indicated by eToro's performance metrics.

Although we do not have access to the signals that were actually displayed to the site's users, we attempt to reconstruct proxies of these signals for our study. For popularity, we use a reconstructed version of the number of copiers each user has based on the transactions in our data. For quality, we use a rough measurement of recent performance: the performance of a user on a particular day is that user's average expected daily return from closed trades in the last 5 business days (more specifically, the average over the subset of those days on which a user had any trading activity).

In this work, we use 100 consecutive days of data, (ranging from September 09, 2011 until December 29, 2011 rather than the first 100 to ensure we have good estimates of popularity), and we reserve the remainder of the data for validation in a planned extension of the current study. This subset includes the trading activity of 24,587 users.

## 4. RESULTS

To support our model and our hypothesis, we approximate $f$ by a linear function of performance. Our hypothesis then reduces to showing that there is a positive significant interaction between past popularity and past performance in a linear regression of change in popularity for each user. To investigate this interaction, we use our measurements of the popularity and performance of each user on each day.

The regressions we examine include data points that consist of the performance of each active user on each day, the popularities of those users on those days, and the popularities of those users on the following days. For the present analysis, we look only at the subset of day-user pairs on which the user did not lose followers. We make this restriction because once a user has more copiers, it is naturally easier for that user to lose copiers, and this effect could alone give the interaction we hope to provide evidence for.

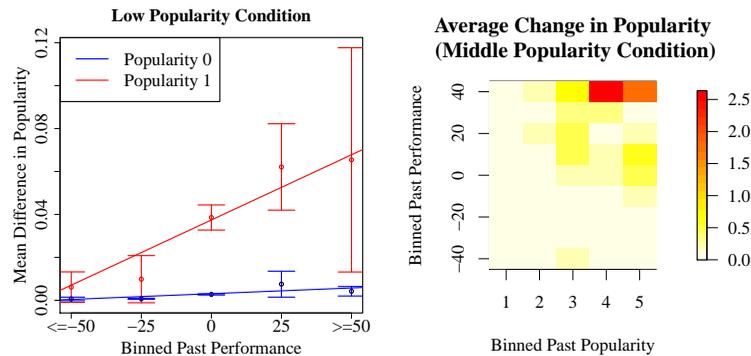

Fig. 1. **Left:** Plot of mean change in popularity against performance (for user-day pairs with popularity zero or one and with 95% confidence Gaussian error bars on the means and regression lines fitted from raw data). To visualize the results of the regression better, we rounded performance to the nearest multiple of 25 and grouped values greater than 50 or less than −50 with those numbers, respectively. These graphical parameters roughly balanced the bin sizes on the x-axis. The difference in slope between the two lines in the third plot displays the hypothesized interaction effect. **Right:** Heat map of mean change in popularity as a function of previous popularity and previous performance for users with greater than zero copiers not in the top 100 most-popular group. For this plot we binned popularities greater than 5 with 5, and we rounded performance to the nearest 10 and grouped values less than −40 or greater than 40 with those numbers, respectively. The nonlinear increase in the third dimensions moving up and to the right in this plot displays the hypothesized interaction.

Since we are also interested in replicating the widely observed cumulative advantage marginal effect of past popularity on future popularity, we also attempt to control for a position bias in the interface of the eToro website. Since users can rank traders by eToro's measurements of performance or by popularity, traders might gain followers just by being displayed prominently on the website. A position bias should not artificially introduce an interaction since users can only sort by one signal at a time, but it could induce artificial marginal effects. To do this control, we further subset the data into two groups: user-day pairs on which the user had 0 or 1 past popularity (i.e., zero copiers or one copier) and user-day pairs on which users had greater than 0 past popularity but on which the user was not among the top 100 most popular users. We use 100 as the cutoff to be conservative. Since viewing the top 100 traders requires scrolling several times, we would not expect the position bias to have an effect in either of these conditions. We attempt to provide evidence for our hypothesis within each of these conditions.

Our hypothesized positive interaction effect and the marginal cumulative advantage effect of popularity are supported in both conditions ($p < 10^{-5}$ for all relevant regression coefficients). Figure 1 displays our results. The plot on the left in Figure 1 shows that having just one copier substantially increases the rate of increase in popularity as a function of performance over having no copiers. The plot on the right shows that a similar trend holds for greater values of popularity as well. We thus conclude that, for this range of popularity values, popularity may amplify the effect of performance in determining the magnitude of increase in future popularity. Perhaps the rich get richer for good reason.

## Acknowledgements

This research was partially sponsored by the Army Research Laboratory under Cooperative Agreement Number W911NF-09-2-0053. Views and conclusions in this document are those of the authors and should not be interpreted as representing the policies, either expressed or implied, of the sponsors. This material is based upon work supported by the National Science Foundation Graduate Research Fellowship under Grant No. 1122374. Any opinion, findings, and conclusions or recommendations expressed in this material are those of the authors(s) and do not necessarily reflect the views of the National Science Foundation.